\documentclass[aps,prl,amsmath,amssymb,superscriptaddress,reprint,floatfix,twocolumn]{revtex4-1}
\usepackage{amssymb, amsbsy, amsmath, latexsym, dsfont, array, layout, graphicx, mathrsfs,xcolor,soul}
\usepackage{natbib}
\usepackage{mathtools}
\usepackage{qcircuit}
\usepackage{physics}
\usepackage[colorlinks,linkcolor=blue,anchorcolor=red,citecolor=red]{hyperref}
\usepackage[capitalize]{cleveref}
\usepackage{todonotes}
\usepackage{textcomp}
\usepackage{bm}
\usepackage{cleveref}

\begin{document}

\title{Experimental device-independent certification of indefinite causal order}

\author{Dengke Qu}
\author{Quan Lin}
\author{Lei Xiao}
\author{Xiang Zhan}
\affiliation{School of Physics, Southeast University, Nanjing 211189, China}
\affiliation{Key Laboratory of Quantum Materials and Devices of Ministry of Education, Southeast University, Nanjing 211189, China}
\author{Peng Xue}\email{gnep.eux@gmail.com}
\affiliation{School of Physics, Southeast University, Nanjing 211189, China}
\affiliation{Beijing Computational Science Research Center, Beijing 100084, China}

\begin{abstract}
Understanding the physical world fundamentally relies on the assumption that events are temporally ordered, with past events serving as causes for future ones. However, quantum mechanics permits events to occur in a superposition of causal orders, providing new types of quantum resources for quantum information tasks. Previous demonstrations of indefinite causal order have relied on a process known as quantum switch and depended on specific assumptions about the devices used in the laboratory. Recently, a theoretical scheme for the certification of indefinite causal order in the quantum switch has been obtained solely from the output statistics of the devices, analogous to the device-independent proofs of nonlocality through violations of the Bell inequality.
Here, we report an experimental verification of the causal inequality using spacelike-separated entangled photons, where one photon functions as the control qubit in a quantum switch and the other serves as an additional observer. Through local measurement statistics, we observe a violation of the causal inequality by $24$ standard deviations.
This work provides evidence for a device-independent certification of indefinite causal order, relying solely on observed correlations without requiring device characterization. Our results pave the way toward a complete understanding of indefinite causal order and its potential applications in quantum information processing.
\end{abstract}
\maketitle

{\it Introduction---}Quantum theory allows the order between events to be coherently controlled by a quantum system~\cite{Hardy07,Chiribella2013Aug,Oreshkov2012Oct,Brukner2014Apr,ABC+15}.
This procedure, known as the quantum switch, has been the subject of much theoretical and experimental study over the past decade, revealing numerous advantages in quantum information tasks, including quantum communication~\cite{ESC18,GHH+20}, quantum channel discrimination~\cite{Chiribella2012Oct}, quantum query complexity~\cite{Colnaghi2012Oct,Araujo2014Dec,Renner2022Jun}, quantum communication complexity~\cite{Guerin2016Sep,WTZ+19}, quantum metrology~\cite{Zhao2020May,CB2021Mar,Frey2019Apr}, quantum thermodynamics~\cite{Felce2020Aug,Guha2020Sep,Simonov2022Mar,ZCH+23,NZH+22,CWJ+22}, and more~\cite{CLM+23,SSD+23}.
The quantum switch exhibits a phenomenon known as \emph{indefinite causal order}~\cite{RSH+24,Baumeler2016Jan}. Various methods have been proposed to certify the indefinite causal order, including causal witnesses~\cite{ABC+15,Bavaresco2021Nov,Abbott2017Dec,Abb+16,Wechs2019Jan}, process tomography~\cite{Antesberger2024Feb}, and semi-device-independent approaches~\cite{Bavaresco2019Aug,Dourdent2022Aug}. Many of these certification schemes have also been implemented experimentally~\cite{Rubino2017Mar,Goswami2018Aug,Stromberg2023Aug,Cao2023May,Rubino2022Jan}.

For a long time, a fully device- and theory-independent method for certifying indefinite causal order in the quantum switch, based solely on the observed correlations between measurement settings and outcomes, remained unavailable. The usual approach for device-independent certification of indefinite causal order is based on the violation of \emph{causal inequalities}~\cite{Oreshkov2012Oct}, which is however shown impossible to achieve by using the quantum switch, or indeed any of the more general class of quantum processes with coherent control of causal order~\cite{AFN+17,BAF+16,ABC+15,OG16,Purves2021Sep,Wechs2021Aug}---the only indefinite causal order processes so far studied in the laboratory.

Recently, however, a number of theoretical works have shown that testing indefinite causal order device-independently in the quantum switch is possible in extended scenarios including a spacelike-separated measurement~\cite{vanderLugt2023Sep, vdLO24, GP23, Dourdent2024Oct}. One of these is the work of Van der Lugt, Barrett, and Chiribella (VBC)~\cite{vanderLugt2023Sep}, which shows that the quantum switch violates a Bell-like inequality, derived from assumptions called definite causal order, relativistic causality, and free interventions. The violation of the causal inequality thus demonstrates the existence of indefinite causal order, subject to the assumptions of relativistic causality and free interventions. Experimentally, violating the causal inequality presents a significant technical challenge. It requires high-quality entanglement between the control qubit of the quantum switch and the additional qubit, fast switching of the measurement settings for both parties, and the preservation of coherence between the two causal orders.

In this work, we employ spacelike-separated entangled photon pairs to experimentally investigate the violation of the Bell-like inequality. One of the entangled photons functions as the control qubit of the quantum switch, while the other serves as an additional qubit. We close the locality loophole by ensuring that the local measurement performed by one party is spacelike-separated from the local measurement performed by the other~\cite{Hensen2015Oct,Giustina2015Dec,Shalm2015Dec}. In our experiment, this is achieved by independently and rapidly selecting the measurement settings for both parties such that no physical signal, constrained by the speed of light, can transmit information about the chosen setup or measurement outcome to the other party in time. Both parties randomly select their measurement settings in each experimental run, thereby closing the freedom-of-choice loophole. Furthermore, we adopt the fair-sampling assumption that the sample of measured photons fairly represents the entire ensemble~\cite{CHSH69,CS78}.
Overall, our experiments provide evidence for device-independent verification of indefinite causal order, showing that quantum correlations defy explanation by any hidden variable theory based on a definite causal structure.

\begin{figure}
	\centering
	\includegraphics[width=0.4\textwidth]{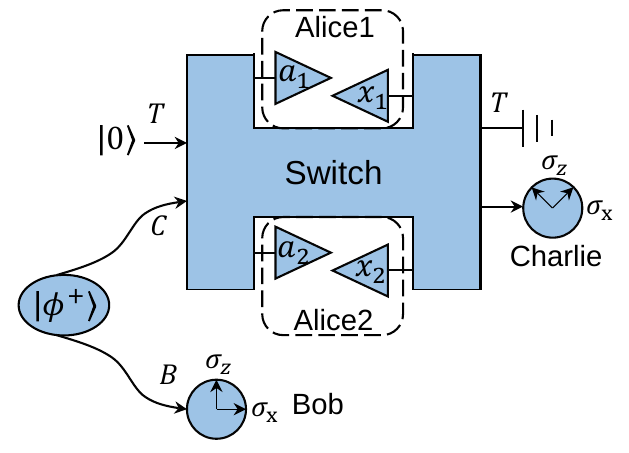}
 	\caption{The quantum switch setup violates the VBC inequality. The control qubit \( C \) and the remote qubit \( B \), located with Bob, are initially prepared in the maximally entangled Bell state \( \ket{\phi^+} \). The target system \( T \) is initialized in the state \( \ket{0} \) and is subsequently measured and re-prepared in the computational basis by Alice~1 and Alice~2 within the switch. After this sequence, \( T \) is discarded. Given their respective inputs \( x_1 \) and \( x_2 \), Alice~1 and Alice~2 obtain outputs \( a_1 \) and \( a_2 \), respectively. Bob and Charlie then perform projective measurements on qubit \( B \) and the final control qubit \( C \), corresponding to their settings \( y \) and \( z \). These measurements are performed along directions confined to the \( XZ \)-plane of the Bloch sphere, represented by black arrows, and yield outcomes \( b \) and \( c \), respectively.}
	\label{fig:1}
\end{figure}

{\it Theoretical background---}As shown in Fig.~\ref{fig:1}, the device-independent scenario considered in Ref.~\cite{vanderLugt2023Sep} consists of four parties, each of whom has a binary measurement setting and binary measurement outcome: Alice 1 (with setting $x_1$ and outcome $a_1$), Alice 2 ($x_2$ and $a_2$), Bob ($y$ and $b$), and Charlie ($z$ and $c$).
Repeatable experiments by these four parties thus produce probability distributions of the form of $p(a_1 a_2 b c | x_1 x_2 y z)$.
\emph{Causal inequalities}, from literatures preceding Ref.~\cite{vanderLugt2023Sep}, are derived from assumptions of freely chosen interventions and the existence of a definite (though possibly \emph{dynamic}) causal order between all parties~\cite{OG16, Abb+16}.
These causal inequalities have been shown not to be violated by the quantum switch~\cite{ABC+15,AFN+17,BAF+16,OG16}.
What sets apart the derivation of the inequality in Ref.~\cite{vanderLugt2023Sep} is that an additional assumption is made---called relativistic causality---which constrains the allowed causal orders.
More precisely, Bob is placed at spacelike separation from the three other parties, while Charlie is in the future lightcone of Alice 1 and Alice 2.
The assumption of relativistic causality constrains the allowed causal orders to those consistent with this spacetime setup, leaving only the order between Alice 1 and Alice 2 open and subject to possible indefiniteness.

The mathematical formulations of the assumptions of definite causal order, relativistic causality, and free interventions jointly lead to the inequality
\begin{align}\label{eq:the-inequality}
&p\left(b=0, a_2=x_1 \mid y=0\right)+p\left(b=1, a_1=x_2 \mid y=0\right)\\ \nonumber
&+p\left(b \oplus c=y z \mid x_1=x_2=0\right) \leq \frac{7}{4}.
\end{align}
In the following, we refer to this as the \emph{VBC inequality}. Its derivation is based on techniques from the study of Bell inequalities; however, crucially, it does not assume Bell locality.

Quantum-theoretically, the constrained causal inequality can be violated by a setup involving a quantum switch, depicted in Fig.~\ref{fig:1}.
The quantum switch has a \emph{target qubit} $T$ as well as a \emph{control qubit} $C$.
Initially, the target qubit \( T \) is prepared in the state \( \ket{0}_T \), while the control qubit \( C \) is maximally entangled with a distant qubit \( B \) in the Bell state \( \ket{\phi^+} = (\ket{00}_{BC} + \ket{11}_{BC})/\sqrt{2} \).
The quantum switch, as defined in Ref.~\cite{Chiribella2013Aug}, enables Alice~1 and Alice~2 to act on the target qubit in an order that is coherently controlled by the state of the control qubit.
Each party, Alice~\( i \) (with \( i = 1,2 \)), performs a measurement on the incoming target qubit in the computational basis \( \{ \ket{0}_T, \ket{1}_T \} \), irrespective of the input \( x_i \), and record the outcome as $a_i$. She then re-prepares the target qubit in the state \( \ket{x_i}_T \).
Once both operations are completed, the target qubit is discarded. Charlie subsequently measures the output control qubit.
He measures in the $\sigma_z+\sigma_x$ direction if his setting $z=0$ and in the $\sigma_z-\sigma_x$ direction if $z = 1$, recording his outcome as the variable $c$.
Meanwhile, Bob, who possesses the spacelike-separated qubit \( B \), carries out a measurement in the \( \sigma_z \) basis if \( y = 0 \), or along \( \sigma_x \) if \( y = 1 \), and notes his outcome as \( b \).
Note that Bob and Charlie are thus in essence performing a Bell-type experiment on the shared entangled state on $BC$, with the crucial difference that qubit $C$ is first used as the control qubit of a quantum switch.

Following this protocol, quantum theory predicts values of $1/2$, $1/2$, and $1/2 + \sqrt 2/4$ for the three respective terms in \eqref{eq:the-inequality}, adding up to
\begin{equation}
    3/2 + \sqrt2/4 \approx 1.8536 > 7/4,
\end{equation}
violating the constrained causal inequality \eqref{eq:the-inequality}---and thus demonstrating that at least one of the assumptions of definite causal order, relativistic causality, and free interventions is violated.

\begin{figure}
	\centering
	\includegraphics[width=0.45\textwidth]{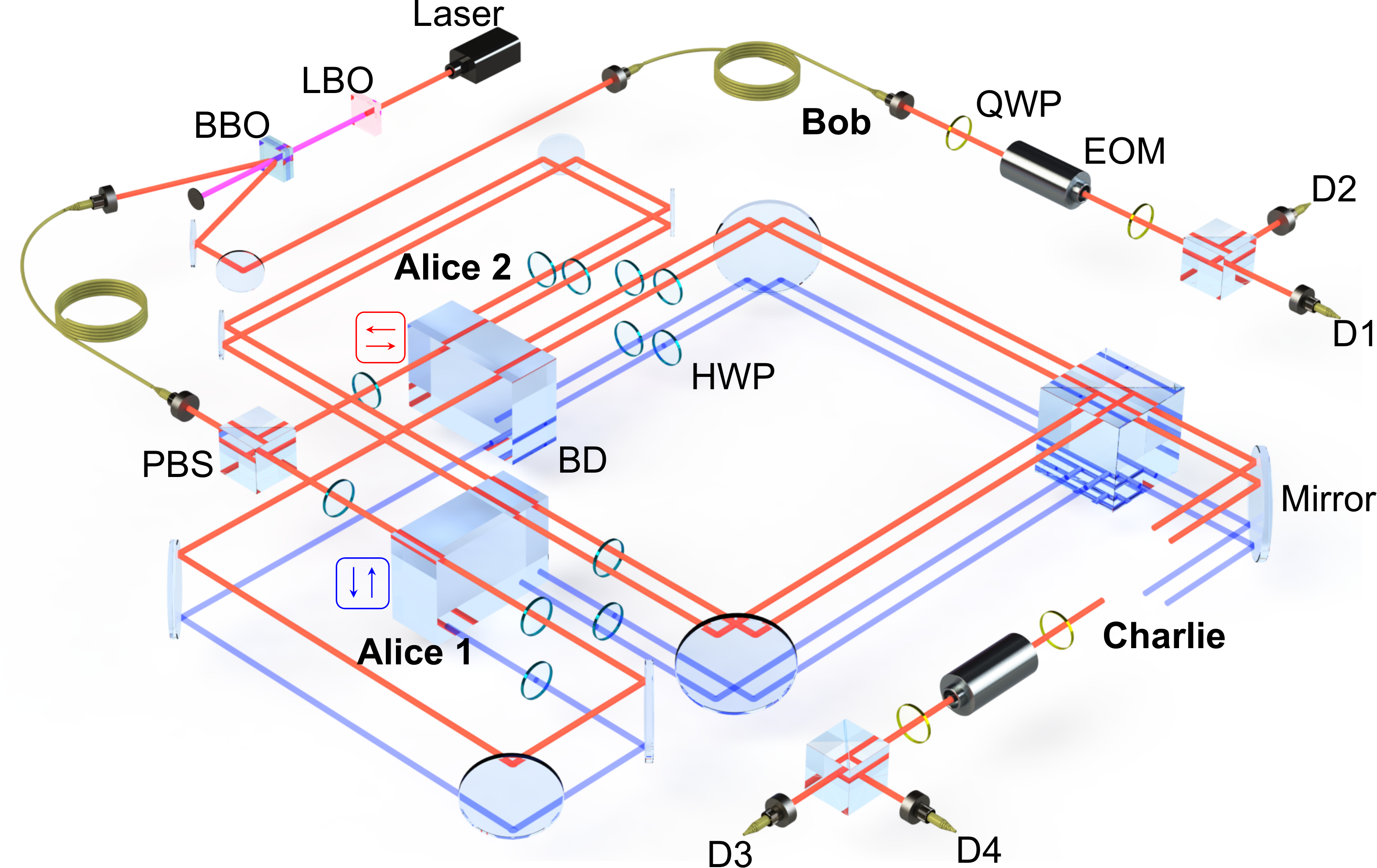}
 	\caption{Experimental setup of the quantum switch. A femtosecond laser pulse at $780$ nm is frequency-doubled in a lithium triborate (LBO) crystal to produce a $390$ nm pump beam for two beam-like cut $\beta$-barium borate (BBO) crystals, generating entangled photon pairs via type-II spontaneous parametric down-conversion. One photon is directed to Bob for immediate measurement, while its partner is routed through the quantum switch and then measured by Charlie. Measurements employ an electro-optic modulator (EOM), two quarter-wave plates (QWPs), and a polarization beam splitter (PBS). Inside the switch, the photon’s polarization and path modes serve as the target and control qubits, respectively. Alice $1$ and Alice $2$ perform measurement-and-reprepare operations on the target qubit using a beam displacer (BD) and half-wave plates (HWPs). The four interferometer loops form a \(2\times2\) optical-path array, corresponding to the combined outcomes of Alice $1$ and Alice $2$. Finally, photons are detected by avalanche photodiodes (APDs).}
	\label{fig:2}
\end{figure}

{\it Experimental implementation---}To test the device‐independent causal inequality, we implement a photonic quantum switch by employing a pair of spacelike‐separated entangled photons. We encode the basis of the qubit as $\ket{0}=\ket{H}, \ket{1}=\ket{V}$, where $\ket{H}~(\ket{V})$ denotes horizontal (vertical) polarizations of the photons.
As illustrated in Fig.~\ref{fig:2}, two closely spaced, $2$ mm-thick, type-II phase-matched $\beta$-barium borate (BBO) crystals, which are separated by a true zero-order half-wave plate (HWP), are pumped by a $390$ nm laser to generate the maximally entangled state
\begin{equation}
\ket{\phi^+}=\frac{1}{\sqrt{2}}(\ket{HH}+\ket{VV}).
\end{equation}
The coincidence count rate of the entangled photon pairs is $6000~\text{s}^{-1}$.
The fidelity of the maximally entangled state is $0.9884\pm0.0018$.

\begin{figure*}
	\centering
	\includegraphics[width=0.95\textwidth]{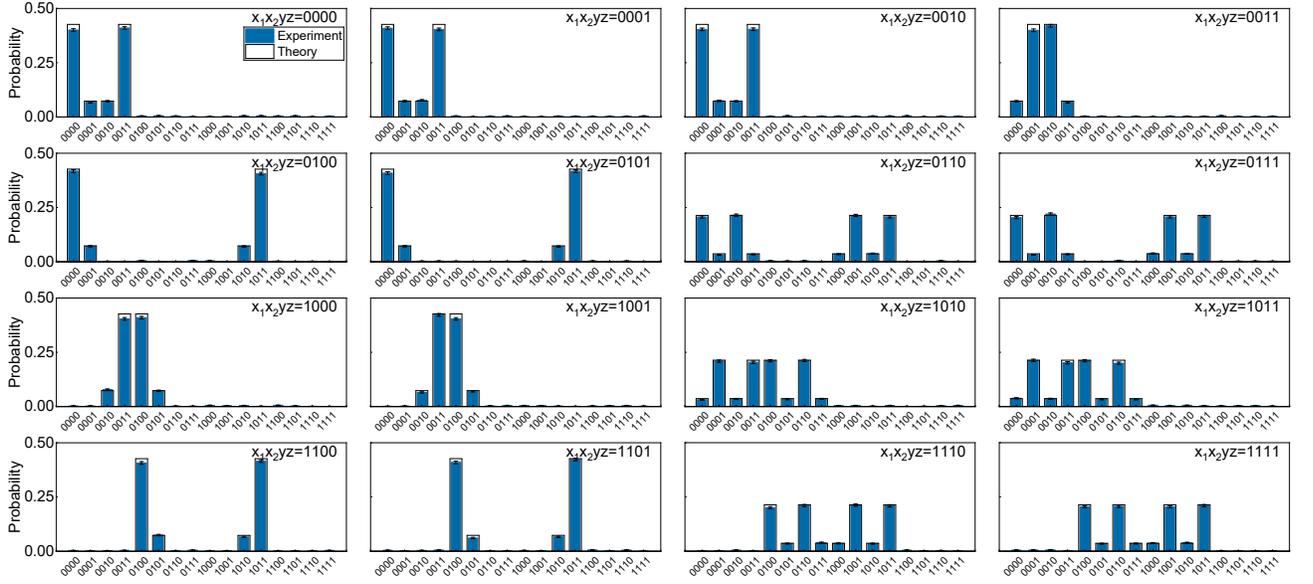}
 	\caption{Experimental results of the setting–outcome correlations generated by the quantum switch. Each panel displays the conditional probability distribution of the joint outcomes $\{a_1,a_2,b,c\}$ given an input setting $\{x_1,x_2,y,z\}$. The blue bars indicate the experimental results, while the hollow bars represent the theoretical predictions. The experimental uncertainties arise from photon-counting statistics and correspond to the standard deviations.}
	\label{fig:3}
\end{figure*}

One photon of the entangled pair propagates through $20$ m of single‐mode fiber to Charlie, located $15$ m from the source, where it functions as the control qubit of the quantum switch.
To implement the quantum switch, we map the polarization of the photon onto spatial modes to encode the control qubit and encode the target qubit in the polarization.
Specifically, horizontally polarized photons $\ket{H}$ are transmitted by the PBS into the right path $\ket{R}$, where Alice $1$ performs perform measurement and re-prepare instruments before Alice $2$.
In contrast, vertically polarized photons $\ket{V}$ are reflected by the PBS into the upper path $\ket{U}$, where Alice $2$ performs the operation before Alice $1$.
To initialize the target qubit in the state $\ket{0}$, half-wave plates (HWPs) are inserted in each spatial mode: one set to $45^\circ$ in path $\ket{R}$ and another set to $0^\circ$ in path $\ket{U}$, ensuring that photons in both modes emerge with horizontal polarization $\ket{H}$. The initial state of the whole system is $\ket{\psi_0}=\frac{1}{\sqrt{2}}(\ket{HR}_{BC}+\ket{VU}_{BC})\otimes\ket{H}_T$, where the subscripts represent the various subsystems.

Alice$~1$ and Alice$~2$ apply their respective instruments to the target qubit. Each measurement employs a beam displacer (BD) to separate the photon into distinct spatial modes according to the outcome \(a_i\). The deflecting direction of the BDs for Alice$~1$ and Alice$~2$ are orthogonal.
On Alice$~1$’s side, horizontally polarized photons remain in the original mode (\(a_1 = 0\)), while vertically polarized photons are deflected downward (\(a_1 = 1\)). In contrast, on Alice$~2$’s side, vertically polarized photons remain undeviated (\(a_2 = 1\)), and horizontally polarized photons are deflected to the left  (\(a_2 = 0\)).
The joint outcomes \((a_1, a_2)\) define a \(2\times 2\) interference array, with the upper and lower paths indicated by red and blue beams in the Fig.~\ref{fig:2}, respectively. Following the BDs, HWPs set to either $0^\circ$ or $45^\circ$ are inserted according to the setting \(x_i\) of each party, rotating the target qubit into a single-polarization state.
In this process, Alice~1 and Alice~2 perform local operations on the target system, respectively, but delay reading out their outcomes until the information about the order of their operations has been erased, thereby preserving a coherent superposition of causal orders.
Before completing the quantum switch, the HWPs set to appropriate angles (not shown in Fig.~\ref{fig:2}) and following a PBS recombine the interference loops with identical outcomes \((a_1,a_2)\) into a single path, which erases the target-qubit information and transforms the control qubit from path encoding back to polarization encoding ($\ket{R}\to \ket{H}$, $\ket{U} \to \ket{V}$).

In the next stage, Charlie performs the measurement on the control qubit.
The measurement is selected by a binary sequence generated by a random number generator (RNG): the control signal 0 triggers the  measurement of the observable $\sigma_z+\sigma_x$ and 1 triggers the measurement $\sigma_z-\sigma_x$.
These observables are represented as \( M = \sum_{i=H,V} m_i \ketbra{m_i} \), where \( \ket{m_i} \) are the eigenstates and \( m_i \) the associated eigenvalues. To realize the measurement, a polarization rotation operator \( U_M = \ket{H}\bra{m_H} + \ket{V}\bra{m_V} \) is applied to the photon's polarization.

This unitary transformation is implemented using a composite optical setup consisting of a quarter-wave plate (QWP) set at \( 0^\circ \), followed by an electro-optic modulator (EOM) with rotation angle \( 2\vartheta \), and a second QWP at \( 90^\circ \).
The EOM contains a birefringent crystal oriented at \( 45^\circ \) relative to the $x/y$ axes, such that its effect on the polarization state can be described by the matrix
$R_\text{EOM}(\vartheta)=\begin{pmatrix}
\cos\frac{\vartheta}{2} & i\sin\frac{\vartheta}{2} \\
i\sin\frac{\vartheta}{2} & \cos\frac{\vartheta}{2}
\end{pmatrix}$. Together with the QWPs, the full polarization rotation operator is given by
$U_M=\begin{pmatrix}
1 & 0 \\
0 & -i
\end{pmatrix}R_\text{EOM}(2\vartheta)\begin{pmatrix}
1 & 0 \\
0 & i
\end{pmatrix}=\begin{pmatrix}
 \cos\vartheta & -\sin\vartheta \\
\sin\vartheta & \cos\vartheta
\end{pmatrix}$.
The projective measurement is implemented after passing through a PBS.
The outcome $c$ corresponds to the measurement result of the photon in the basis state $|i\rangle \in\{|H\rangle,|V\rangle\}$. The interval between the RNG and the measurement is approximately $22.6$~ns, corresponding to a spatial separation of $6.78$ m (see Supplementary Material~\cite{supp} for more details). Photon detection is achieved with avalanche photodiodes (APDs).


The other photon travels through $20$ m of single‐mode fiber to Bob, who is located $9$ m from the source. The photon undergoes the random setting choices and the projective measurement. Similarly, Bob measures the qubit with the observable $\sigma_z$ when $y = 0$ and $\sigma_x$ when $y = 1$. The measurement setting is determined by a binary sequence generated by a RNG. The measurement apparatus consists of two QWPs and an EOM controlled by the RNG followed by a PBS. Bob and Charlie are placed at distinct optical tables, with a physical distance of $20$ m between them. In our experiment, we rapidly and independently select the measurement settings and perform measurements on both sides, guaranteeing that no physical signal—limited by the speed of light—can carry the setting choice or result of Bob to Charlie in time~\cite{supp}, which closes the locality loophole~\cite{Hensen2015Oct,Giustina2015Dec,Shalm2015Dec}.

{\it Experimental result of the causal inequality---}To evaluate the causal inequality, we measure conditional probability distributions $p(a_1,a_2,b,c|x_1,x_2,y,z)$ for all joint settings (Fig.~\ref{fig:3}). The hollow bars indicate theoretical predictions, while the blue bars show experimental probabilities. The experimental results exhibit excellent agreement with the theoretical predictions. The three terms of the VBC inequality~\eqref{eq:the-inequality} can be written as
\begin{align}
&p\left(b=0, a_2=x_1 \mid y=0\right)\\ \nonumber
&=\frac{1}{8} \sum_{x_1, x_2, z \in\{0,1\}} p\left(b=0, a_2=x_1 \mid x_1 x_2 z, y=0\right),\\ \nonumber
&p\left(b=1, a_1=x_2 \mid y=0\right)\\ \nonumber
&=\frac{1}{8} \sum_{x_1, x_2, z \in\{0,1\}} p\left(b=1, a_1=x_2 \mid x_1 x_2 z, y=0\right),\\ \nonumber
&p(b \oplus c=y z\mid x_1=x_2=0)\\ \nonumber
&=\frac{1}{4} \sum_{y, z \in\{0,1\}} p(b \oplus c=y z \mid y z,x_1=x_2=0).
\end{align}

For the first two terms of the inequality, we set the measurement setting $y = 0$, which corresponds to a computational basis measurement performed by Bob. We repeat the procedure over all combinations of measurement settings for Alice$~i~(i=1,2)$. To trace over Charlie's system, we perform measurements in the $\sigma_z + \sigma_x$ and $\sigma_z - \sigma_x$ basis and sum the corresponding results. As shown in Fig.~\ref{fig:3}, the probability distribution of the first two columns is used to calculate the first two terms of the VBC inequality~\eqref{eq:the-inequality}.
For the third term of the inequality, Alice~1 and Alice~2 perform identity operations, ensuring that the target qubit remains unaffected. In this case, the operations performed by Bob and Charlie are equivalent to those in a standard Bell test. The setting-outcome correlations from the first row in Fig.~\ref{fig:3} are used to calculate the experimental value of the third term.

The experimental results for the three terms of the causal inequality are $0.4846\pm 0.0022$, $0.4895\pm 0.0022$, and $0.8348\pm 0.0023$, respectively, in close agreement with the theoretical predictions of $1/2$, $1/2$, and $1/2+\sqrt{2}/4$. Consequently, the measured inequality value is $1.8090\pm 0.0024$, which exceeds the classical bound of $1.75$ by $24$ standard deviations.
All measurement settings are chosen independently, and their spacetime relations follow Fig.~\ref{fig:1}, satisfying the assumptions of definite causal order, relativistic causality, and free interventions.
The experiment results demonstrate that the observed correlations cannot be explained by any fixed causal order model, thereby supporting the existence of indefinite causal order.

{\it Conclusion and discussion---}
In this work, we demonstrate a device-independent violation of a Bell-like inequality using a spacelike-separated entangled source and a photonic quantum switch. The violation of this Bell-like (VBC) inequality implies that it is not possible to assign hidden variables to a fixed causal order within the quantum switch.
In particular, from a theoretical point of view, discussions of device-independent indefinite causal order often treat Alice$~i$'s event as a combination of the choosing of her setting and the producing of her outcome, which are considered to happen inside her laboratory in short succession (at least as experienced by her). However, in practice, experimental implementations require the settings of Alice~1 and Alice~2 to be selected relatively early, while their outcomes must be read out at a later time. This temporal structure is necessary to prevent the causal order from decohering, and it is a constraint shared by all foreseeable experimental realizations.

Our experiment provides evidence for the existence of an indefinite causal order within a spacetime structure. The violation of the VBC inequality represents a significant step toward a loophole-free verification of indefinite causal order, highlighting its physical reality. Furthermore, the certification of indefinite causal order offers potential for both foundational studies and practical applications in quantum information processing.

\begin{acknowledgments}
{\it Acknowledgments---}This work has been supported by the National Key R$\&$D Program of China (Grant No. 2023YFA1406701), the National Natural Science Foundation of China (Grant Nos. 12025401, 92265209, 12305008 and 92476106), the National Postdoctoral Program for Innovative Talent (Grant Nos. BX20230036 and BX20240065), the China Postdoctoral Science Foundation (Grant Nos. 2023M730198 and 2024M750405), and Beijing National Laboratory for Condensed Matter Physics (No. 2024BNLCMPKF010).
We thank Tein van der Lugt for helpful discussions.
\end{acknowledgments}

{\it Data availability---} The data supporting this study's findings are available within the Letter~\cite{supp2}.

{\it Note added---}During the completion of our manuscript, we became aware of two works by Richter et al.~\cite{Richter2025} and Guo et al.~\cite{guo2025}, which independently demonstrated the protocol.

\end{document}